# NIME: A Community of Communities

Michael J. Lyons

Ritsumeikan University

lyons@im.ritsumei.ac.jp

Commentary on "Fourteen Years of NIME: The Value and Meaning of 'Community' in Interactive Music Research" by A. Marquez-Borbon & P. Stapleton

I am not sure when I first heard the phrase 'the NIME community'. The current community mailing list, community@nime.org, named in 2007, might suggest NIME has consciously considered itself as a community for nearly a decade. However, I have the impression that only in recent years has NIME been frequently described as a community, both verbally and in articles published at the conference.

The current article, the most recent in this collection, was selected because it is the first to explicitly analyze NIME community. The authors base their analysis on the theoretical frameworks of 'situated learning' (SL) and 'communities of practice' (CoP), which have long enjoyed broad influence in fields related to the sociology of learning, knowledge, and innovation. It may be helpful to read this article in conjunction with some of the literature on SL and CoPs. The lucid and insightful work of Brown and Duguid (1991) is particularly recommended.

The article is, therefore, not really a survey of the NIME community, as the title might lead one to think, but an investigation of the NIME community from a particular theoretical framework. This does, however, offer readers with some experience at NIME a valuable opportunity to reflect on the social processes at work in the community from a perspective they might not normally adopt.

Based on a taxonomy from the CoP literature, the authors conclude that the NIME community is not itself a CoP, because it is too heterogeneous in approach. It more closely resembles a 'community of interest' (CoI), in that it aims broadly to *"develop a body of work related to new digital instruments from different disciplines and perspectives."*

Admirably, the authors do not attempt to force-fit NIME into the CoP framework, writing instead that: "Practitioners from different domains – HCI, computer science, electrical engineering, (computer) music, and arts – engage in a joint enterprise. Perhaps NIME could appropriately be described as a 'community-of-communities'."

This resonates closely with my view. Multi-disciplinarity was an important aspect of the NIME workshop proposal (Poupyrev et al. 2001) and has continued to factor in the conference (Lyons and Fels, 2015), which is open to a plurality of approaches, tastes, and levels of expertise in both music and technology.

The authors subsequently locate bona-fide CoPs in several subset communities affiliated with NIME. This treatment is valuable and could serve as the basis for interesting further work, but it is again not a survey, since there are many such specialized groups, and these have tended to evolve rapidly, sometimes spawning independent events. Moreover, the article does not further analyze features of the community as a whole, and this leads me to ask: What other approaches could shed light on the NIME community?

Besides a theoretical critique, it could be revealing to attempt empirical work. Scientometric analyses of the NIME published proceedings, such as Jensenius' (2014) analysis of gesture-related terminology can also provide insight into aspects of the community. On a larger scale, Liu et al. (2014) presented a detailed and fascinating empirically-based account of the CHI research community using co-word analysis. This not only maps the network of research interests of the community, but highlights the connectivity of interests and how these have changed over time. It should be possible to conduct a similar analysis of the NIME archive and this would serve to map and track the evolution of the activities and interests of the community.

Scientometrics does not directly address the sociological questions here and an ethnographic approach may be necessary. However, rigorous ethnographic studies, such as underlie CoP theory itself, would present special challenges in the context of NIME, a shape-shifting community that gathers for only a fews days of the year, and new musical interface research may be considered too marginal a practice to attract the resources needed for such a large-scale project. Sub-CoPs should be amenable to ethnographic work, suggesting a potentially interesting line of research, but it is important to take care that focussing on sub-groups does not lead to a reductionist view that ignores the ecology of the larger community.

A more modest idea would be to survey the participants themselves about their experience of the NIME community, using interviews or free-form reflection. Such a study could leverage existing resources such as the conference mailing list and web portal. Designed intelligently, this could serve as a participatory, awareness-raising, and potentially community-building activity.

**References**

Brown, John Seely, and Paul Duguid. "Organizational learning and communities-of-practice: Toward a unified view of working, learning, and innovation." *Organization science* 2, no. 1 (1991): 40-57.

Jensenius, A. R. (2014). To gesture or not? An analysis of terminology in NIME proceedings 2001–2013. In *Proceedings of the International Conference on New Interfaces For Musical Expression*, pages 217–220, London.

Liu, Yong, Jorge Goncalves, Denzil Ferreira, Bei Xiao, Simo Hosio, and Vassilis Kostakos. "CHI 1994-2013: Mapping two decades of intellectual progress through co-word analysis." In *Proceedings of the 32nd annual ACM conference on Human factors in computing systems*, pp. 3553-3562. ACM, 2014.

Lyons, Michael J., and Sidney S. Fels. "Introduction to Creating Musical Interfaces." In *Proceedings of the 33rd Annual ACM Conference Extended Abstracts on Human Factors in Computing Systems*, pp. 2481-2482. ACM, 2015.

Poupyrev, Ivan, Michael J. Lyons, Sidney Fels, and Tina Blaine (Bean). "New interfaces for musical expression." In *CHI'01 Extended Abstracts on Human Factors in Computing Systems*, pp. 491-492. ACM, 2001.